\documentclass[%
 aps,
 prb,
 amsmath,amssymb,
 showpacs,
 reprint,%
]{revtex4-2}

\usepackage{graphicx}
\usepackage{dcolumn}
\usepackage{bm}
\usepackage{xcolor} 
\usepackage{verbatim}

\begin{document}

\title{Real-space $d$-wave superconductivity from weak attraction}

\author{Pavel Kornilovitch}
 \email{pavel.kornilovich@gmail.com}
 \affiliation{Department of Physics, Oregon State University, Corvallis, Oregon 97331, USA} 

\date{\today}  

\begin{abstract}

It is shown that even a weak out-of-plane attraction, $V \rightarrow 0$, can form real-space pairs in the body-centered tetragonal lattice despite the presence of a Hubbard repulsion and a fully developed three-dimensional kinetic energy. In the hole channel, the pairs have $d$ orbital symmetry, which, following Bogoliubov's argument, translates into a $d$-symmetric macroscopic superconducting order parameter. The findings help us to understand key features of La$_{2-x}$Sr$_x$CuO$_4$ and further support the real-space mechanism of superconductivity in cuprates. The rules of real-space superconductivity are formulated at the end.    

\end{abstract}



\maketitle

\section{\label{D:sec:one}
Introduction
}

In BCS superconductivity~\cite{Bardeen1957,Schrieffer1999}, Cooper pairs can form because the Fermi sea reduces kinetic energy~\cite{Cooper1956} while slow ions effectively reduce Coulomb repulsion~\cite{Bogoliubov1958,Tolmachev1961,Morel1962}. In BEC superconductivity~\cite{Ogg1946,Schafroth1954a,Schafroth1954b,Schafroth1957,Blatt1964,Eagles1969,Bogoliubov1970,Alexandrov1981,Micnas1990,Alexandrov1994}, real-space pairs can form if a non-retarded attraction $V$ overcomes repulsion $U$ and kinetic energy $K$. Pair formation threshold, $V_{\rm cr}$, is of the order of several hopping integrals $t$ for a broad range of lattice geometries \cite{Kornilovitch2023}. $V_{\rm cr}$ remains finite even in the absence of repulsion as long as $K$ is three dimensional. In hole-doped cuprates near-half filling, $t \approx 0.1$ eV \cite{Harrison2023}, which means a $V_{\rm cr}$ of several tenths of eV is needed to form pairs. This may be challenging to achieve with phonon \cite{Alexandrov1994}, Jahn-Teller~\cite{Bednorz1988,Mihailovic2001,Kabanov2002}, or spin-wave~\cite{Scalapino2012} mediation. One approach to lower $V_{\rm cr}$ is to spread attraction over several unit cells trading the depth of an attractive potential for its width \cite{Kornilovitch1995}. In this paper, we describe a realistic three-dimensional (3D) system where an {\em infinitesimal} out-of-plane attraction $V \ll t$ produces real-space pairs despite the presence of a Hubbard repulsion and fully developed 3D kinetic energy. In the hole channel of the body-centered tetragonal (BCT) lattice, $q_z$-independent dispersion near $q_x , q_y = \pm \pi$ renders hole movement near the band edges effectively two dimensional. A nonlocal $V$ neutralizes Hubbard repulsion $U$. As a result, pairs are formed with zero threshold, $V_{\rm cr} \rightarrow 0$, even for finite $U$ and interlayer hopping $t_{\perp}$. The BCT lattice serves as a prototype structure of La$_{2-x}$Sr$_{x}$CuO$_4$ (LSCO) where each BCT site represents one CuO$_6$ octahedron. Thus formed real-space hole pairs have $d$-orbital symmetry, which naturally leads to macroscopic $d$-wave superconductivity.

\section{\label{D:sec:two}
BCT lattice as a prototype model structure for LSCO
}

We approach cuprate superconductivity as an anisotro\-pic but {\em three-dimensional} phenomenon. Superconductivity is a phase-coherent Bose condensation of anisotropic real-space pairs formed above $T_c$. This viewpoint is supported by the phenomenology of anisotropic Bose gas  applied to cuprates~\cite{Micnas1990,Alexandrov1994,Uemura1989,Uemura1991,Alexandrov1993b,Geshkenbein1997,Alexandrov1999b,Chen2005,Chen2024}, as well as by measurements of magnetic fluctuations~\cite{Tutueanu2023} and of the interplane coherence length~\cite{Mangel2023}. Real-space pairs above $T_c$ have been observed in cuprates~\cite{Zhou2019} and in iron-based superconductors \cite{Seo2019,Kang2020}. 

\begin{figure}[t]
\includegraphics[width=0.48\textwidth]{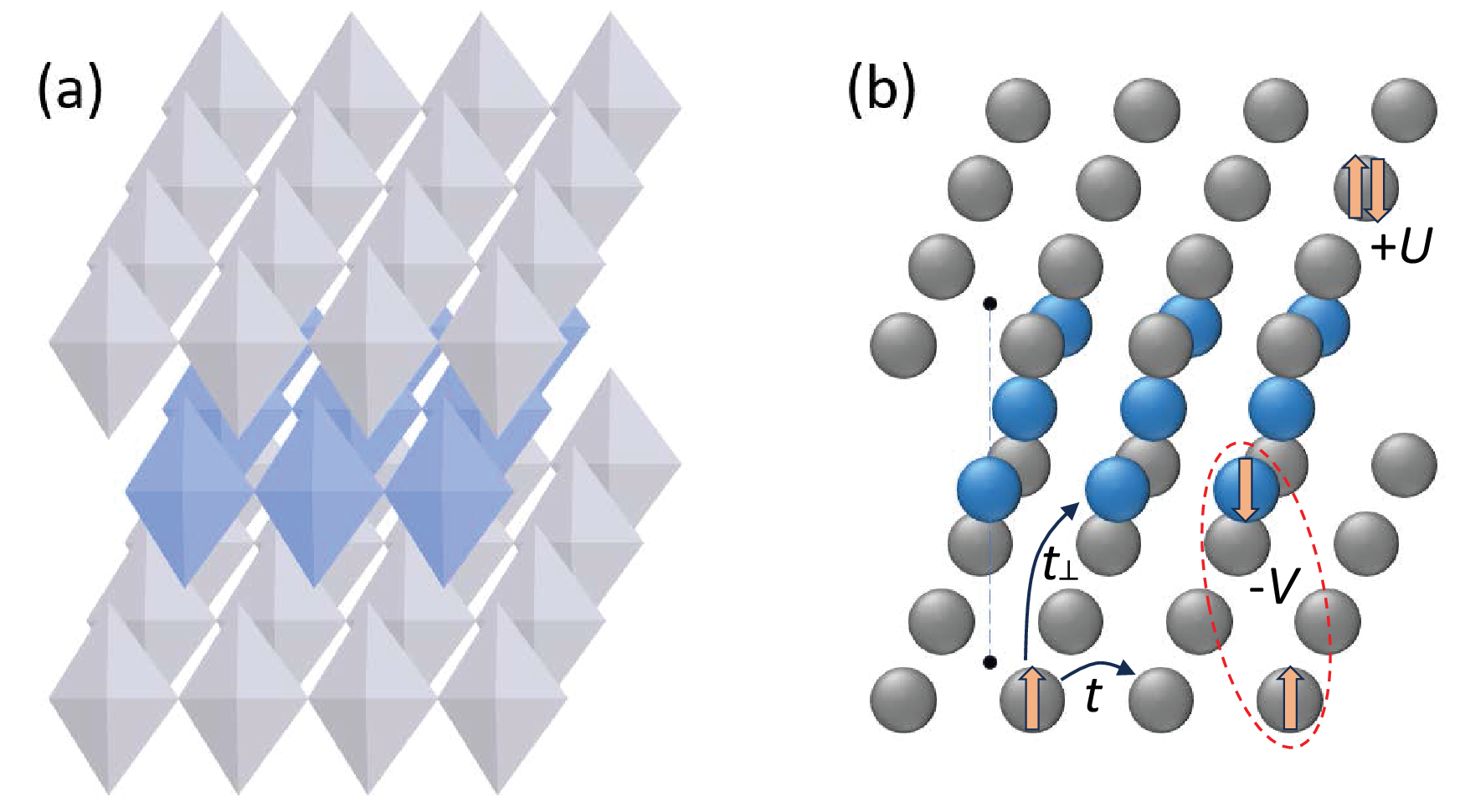}
\caption{(a) CuO$_6$ octahedra in a BCT arrangement. (b) An effective one-orbital BCT lattice. The middle layer is shown in blue for visual purposes.} 
\label{D:fig:one}
\end{figure}

Theoretical description of cuprate superconductivi\-ty should thus be based on three-dimensional models. We focus on the well-studied compound La$_{2-x}$Sr$_x$CuO$_4$. Its main structural elements are CuO$_6$ octahedra arranged in a BCT lattice, see Fig.~\ref{D:fig:one}(a). (The weak orthorhombic distortion that occurs at low temperatures is disregarded here.) In a commonly held view, the most important orbitals are $d_{x^2-y^2}$ of copper ions and $p_x, p_y$ of in-plane oxygen ions~\cite{Emery1987}. We expand this set by including $p_z$ orbitals of apical oxygen ions \cite{Pavarini2001,Barisic2022} that facilitate hole hopping between layers, $t_{\perp}$. We also assume that a multi-orbital model can be down folded to an effective one-orbital model using one of the several available reduction schemes~\cite{Zhang1988,Sakakibara2010,Hirayama2018,Jiang2023}.  An essential feature of such reductions is checkerboard arrangement of orbital phases in the ground state of the effective model, which is a consequence of the odd symmetry of $p_x, p_y$ oxygen orbitals. It results in overall $d_{x^2-y^2}$ symmetry of in-plane holes. In order to preserve this feature in the effective model built of $s$ orbitals, in-plane nearest-neighbor hopping must be positive, $t > 0$. (We leave aside the powerful idea that in-plane hopping can be doping dependent, which may help explain the superconducting dome~\cite{Harrison2023,Kornilovitch2023}.) In physical units, $t \approx 0.1$ eV \cite{Harrison2023}. One should add that both $t$ and $t_{\perp}$ can be affected by electron-phonon interaction and formation of polarons \cite{Alexandrov1996,Kornilovitch1999,Chang2024}.

Inter-hole interaction is discussed next. First, there is a Hubbard repulsion $U \approx 5-6$ eV between two holes occupying the same site \cite{Ramadhan2022,Chainani2023}. Second, quantum-chemical calculations predict a negative excess energy between two {\em static} holes, one on an in-plane oxygen and another on an apical oxygen, of about 0.12 eV \cite{Zhang1991,Catlow1998}. This is interpreted as an out-of-plane attractive potential $V \approx (1 - 2) \, t$. The involvement of apical oxygens in hole pairing is strongly supported by terahertz excitation experiments \cite{Hu2014,Ishioka2023}. A similar-scale near-neighbor attractive potential, $V \approx 1.0 \, t$, was inferred from spectroscopic analysis of 1D cuprates \cite{Chen2021,Wang2021}. In our model, both in-plane and apical oxygens belong to the same effective BCT site. To preserve out-of-plane nature of $V$, we include attraction between two adjacent BCT layers.

We have arrived at the following effective model, see Fig.~\ref{D:fig:one}(b):     
\begin{eqnarray}
H & = &  t \sum_{{\bf m} {\bf b} \sigma} 
c^{\dagger}_{{\bf m} \sigma} c_{{\bf m} + {\bf b}, \sigma} 
       + t_{\perp} \sum_{{\bf m} {\bf b}^{\prime} \sigma} 
c^{\dagger}_{{\bf m} \sigma} c_{{\bf m} + {\bf b}^{\prime}, \sigma}  
\nonumber \\
 &  & + \frac{U}{2} \sum_{\bf m} {\hat n}_{\bf m} ( {\hat n}_{\bf m} - 1 ) 
      - \frac{V}{2} \sum_{{\bf m} {\bf b}^{\prime}} {\hat n}_{\bf m} {\hat n}_{{\bf m} + {\bf b}^{\prime}} \: . 
\label{D:eq:four}
\end{eqnarray}
Here, ${\bf m}$ numbers BCT lattice sites, $\sigma$ is spin projection, and ${\hat n}_{\bf m} = \sum_{\sigma} c^{\dagger}_{{\bf m} \sigma} c_{{\bf m} \sigma}$ is the density operator. ${\bf b}$ and ${\bf b}^{\prime}$ are in-plane and out-of-plane nearest-neighbor vectors, respectively. $t$ and $t_{\perp}$ are in-plane and out-of-plane hopping amplitudes. In this paper, we focus on the hole sector, $t > 0$. We propose Eq.~(\ref{D:eq:four}) as an effective model for underdo\-ped LSCO. The model is consistent with a small Fermi surface on the $x$ side of $x$ vs. $1+x$ reconstruction \cite{Proust2019}, and with the positive Hall sign near half-filling \cite{Khait2023}. The model, Eq.~(\ref{D:eq:four}), possesses several unique properties discussed below.

\begin{figure}[t]
\includegraphics[width=0.40\textwidth]{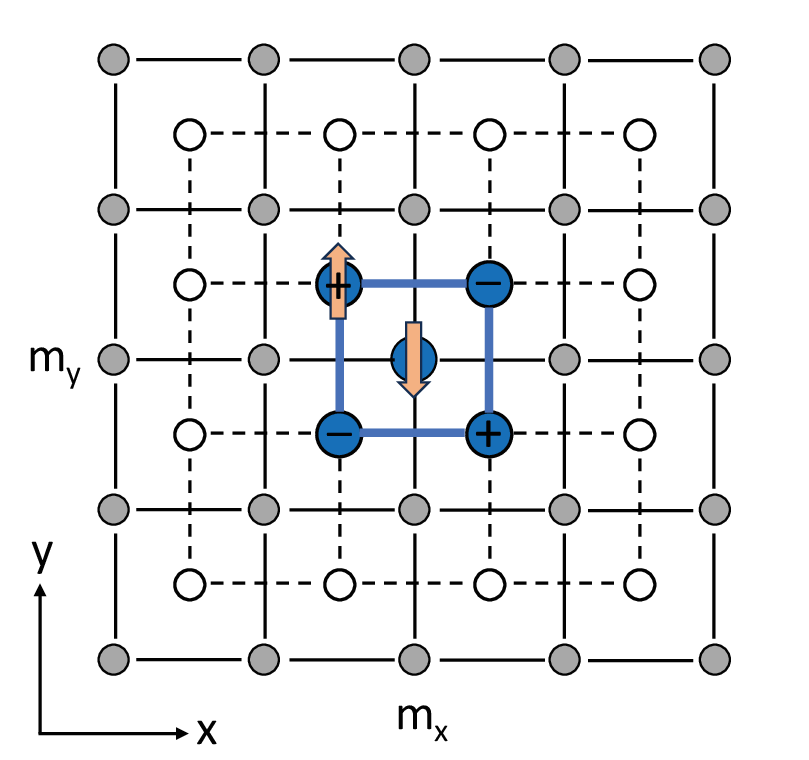}
\caption{Origin of $d$ symmetry in Eq.~(\ref{D:eq:four}). Shown are two BCT layers in top view. The layer of open circles is shifted out of plane (in the $z$ direction) from the layer of filled circles. The hole marked with a down orange arrow is a source of attractive interaction for its pair partner. The latter is confined to the four sites connected by thick blue lines. If $t > 0$, the ground state wave function alternates sign as shown by the pluses and minuses.} 
\label{D:fig:two}
\end{figure}

\section{\label{D:sec:three}
$D$ symmetry
}

$d$ symmetry of the order parameter is a hallmark of cuprate superconductivity~\cite{Tsuei2000} and is one reason the latter is called ``unconventional.'' The experimental situa\-tion, however, remains unclear \cite{Zhu2023}. Sometimes, $d$ sym\-metry is argued to support purely repulsive mechanisms because the node at ${\bf r}_1 = {\bf r}_2$ minimizes Hubbard energy. However, the same argument applies to an extended $s$ wave function that has an even lower energy since it lacks nodes along the angular coordinate. Such reasoning in its simple form is unable to support Hubbard-based mechanisms. Another way to obtain $d$ symmetry within effective models is by introducing a second-neighbor, holelike (positive) hopping in addition to the first-neighbor, electron-like (negative) hopping~\cite{Blaer1997,Bak1999,Singh2022}. $d$-wave superconductivity within the more conventional Eliashberg theory with only electron-phonon interaction has also been reported \cite{Hague2006,Hardy2009,Schrodi2021}.  

Equation~(\ref{D:eq:four}) explains the $d$ symmetry with only nea\-rest-neighbor in-plane hole hopping. The phy\-sics is illustrated in Fig.~\ref{D:fig:two}. Consider relative motion of two holes in the strong-coupling limit, $V \gg t, t_{\perp}$. One pair member resides in an upper BCT layer and is a source of attraction for the second member. The latter is confined within four sites of the lower BCT layer. The problem is isomorphic to a particle on a four-site ring with $t > 0$. The ground state has energy $- V - 2t$ and a wave function that alternates signs as one goes along the ring. It implies a $d$-symmetric pair wave function. This qualitative argument is supported by exact solution of the two-body problem, see Appendix~\ref{D:sec:app:a}, and remains valid in the weak-coupling limit, $V \to 0$. 

We now invoke the relationship between ground-state pair wave function $\psi_0( {\bf r}_1 , {\bf r}_2 )$ and macroscopic order parameter $\Delta( {\bf r}_1 , {\bf r}_2 )$ derived by Bogoliubov for the case of low pair density \cite{Bogoliubov1970} 
\begin{equation}
\Delta( {\bf r}_1 , {\bf r}_2 ) = \sqrt{\frac{N_0}{\Omega}} \, \psi_0( {\bf r}_1 - {\bf r}_2 ) \, , 
\label{D:eq:threeone}
\end{equation}
where $N_0$ is the number of pairs in the condensate and $\Omega$ is the system's volume. For completeness, the derivation is sketched in Appendix~\ref{D:sec:app:b}. According to Eq.~(\ref{D:eq:threeone}), {\em orbital symmetry of the superconducting order parameter is the same as that of the pair wave function}. In other words, a $d$-symmetric hole pair implies a $d$-symmetric order parameter. The effective BCT model, Eq.~(\ref{D:eq:four}), explains $d$-wave superconductivity quite naturally. Physical origin of the $d$ symmetry can be traced back to the positivity of $t$ and, eventually, to the $d_{x^2-y^2}$ and $p_x, p_y$ orbitals of the underlying copper-oxygen plane.

\section{\label{D:sec:four}
Zero binding threshold
}

One-particle dispersion of Eq.~(\ref{D:eq:four}) reads:
\begin{equation}
\varepsilon_{\bf q} = 2 t \, ( \cos{q_x} + \cos{q_y} ) 
+ 8 t_{\perp} \cos{\frac{q_x}{2}} \cos{\frac{q_y}{2}} \cos{\frac{q_z}{2}} \, , 
\label{D:eq:five}
\end{equation}
where we have set BCT lattice parameters, $a = c = 1$. When $t > 0$ and $| t_{\perp} | < \frac{1}{2} \, t $, minimum energy ($\varepsilon_0 = -4t$) occurs at $q_x , q_y = \pm \pi$ and {\em all} $q_z$. Band minimum is not a single point but an entire line. In the ground state, $z$-axis band mass is infinite, and the hole is confined to $(xy)$ plane. Hole motion is effectively two dimensional, and density of states is constant near the band bottom. This is the physical origin of the zero binding threshold.

\begin{figure}[t]
\includegraphics[width=0.48\textwidth]{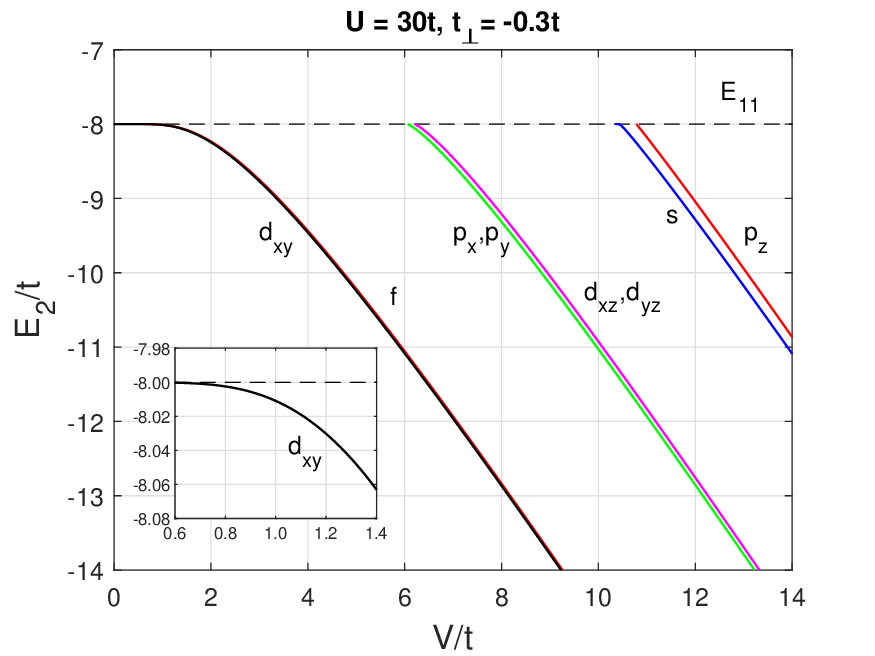}
\caption{Pair energies $E_2$ vs. out-of-plane attraction $V$ in the BCT model, Eq.~(\ref{D:eq:four}), for $U = 30\,t$ and $t_{\perp} = -0.3\,t$. The dispersion equations that define $E_{2}(V)$ are listed in Appendix~\ref{D:sec:app:a}. Notice how the eight states form three groups with close energies. $E_{11} = - 8t$ is the lowest energy of two unpaired holes. Pair formation thresholds for these $U$ and $t_{\perp}$ are: $V_{p_x,p_y} = 6.066t$, $V_{d_{xz},d_{yz}} = 6.211t$, $V_{s} = 10.350t$, $V_{p_z} = 10.786t$. $d_{xy}$ and $f$ pair states have zero formation thresholds. Inset: energy of the lowest pair state, $d_{xy}$, at small $V$.} 
\label{D:fig:three}
\end{figure}

The two-fermion sector of Eq.~(\ref{D:eq:four}) is exactly solvable by methods developed in Ref.~\cite{Kornilovitch2023}. Full analysis is a subject for future research but key pointers are given in Appendix~\ref{D:sec:app:a}. Main findings are summarized here. We assume a positive Hubbard potential, $U > 0$, and limit consideration to $\Gamma$ point of the pair Brillouin zone, i.e., zero total pair momentum. (i) At $V \gg t, t_{\perp}$, there are eight bound pair states. In the spin-singlet sector, they are: extended $s$, degenerate doublet $(d_{xz},d_{yz})$, and a separate $d_{xy}$. In the spin-triplet sector, they are:  degenerate doublet $(p_{x},p_{y})$ and two separate states: $p_z$ and $f$. Only the extended $s$ depends on $U$ while the rest are $U$ independent. (ii) The eight states combine into three groups with close energies, depending on the number of sign changes under $z$ rotations: the low-energy group $(d_{xy},f)$, the mid-energy group $(p_x,p_y,d_{xz},d_{yz})$, and the high-energy group $(s,p_z)$. (iii) The top two groups have finite thresholds that depend on the ratio $t_{\perp}/t$. As $V$ decreases, the top six states disappear into the free-particle continuum. Typical pair energies are shown in Fig.~\ref{D:fig:three}. (iv) Both low-energy states have $d_{xy}$ symmetry under $z$ rotations. Additionally, $f$ state is odd under $(xy)$ plane reflections while $d_{xy}$ is even. Out of the two states, $d_{xy}$ has a slightly lower energy defined by equation     
\begin{equation}
1 = \frac{V}{2} 
   \! \int\limits^{\pi}_{-\pi} \!\! \int\limits^{\pi}_{-\pi} 
 \!\! \int\limits^{2\pi}_{-2\pi} \! \frac{ {\rm d}^3 {\bf q} }{ ( 2 \pi )^3 }
\frac{ ( 1 - \cos{q_x} ) ( 1 - \cos{q_y} ) ( 1 + \cos{q_z} )} 
     { | E_{d_{xy}} | + 2 \varepsilon_{\bf q} } \, . 
\label{D:eq:six}
\end{equation}
When $E_{d_{xy}} \to - 4t - 0$, the integral diverges logarithmically near $q_x , q_y = \pm \pi$, signaling a zero $V$ threshold.

This property is very unusual. Usually, three-dimen\-sional systems require finite attraction to form a bound state. The BCT lattice may be predisposed to low binding thresholds and to real-space superconductivity in general. This finding may contain hints for better understanding of superconductivity in LSCO. The model is also self-consistent. Using the Zhang and Catlow value of $V = 0.12$ eV~\cite{Zhang1991,Catlow1998} and assuming, for example, $t = 600$ K, one obtains from Fig.~\ref{D:fig:three} a binding energy of 210 K, which is the right order of magnitude for the pseudogap in underdoped LSCO~\cite{Tsuei2000,Proust2019}. 

Zero binding threshold should be expected for any three-dimensional dispersion that is $q_z$ independent near the band minimum. Note that this mechanism is different from any enhancement of pairing caused by a van Hove singularity in the middle of the band. Such a singularity is always integrable and by itself cannot lead to a divergent integral. In the present mechanism, it is a constant density of states, rather than the usual square root, near the band's bottom edge causes the entire integral in Eq.~(\ref{D:eq:six}) to diverge.

\begin{figure*}[t]
\includegraphics[width=0.98\textwidth]{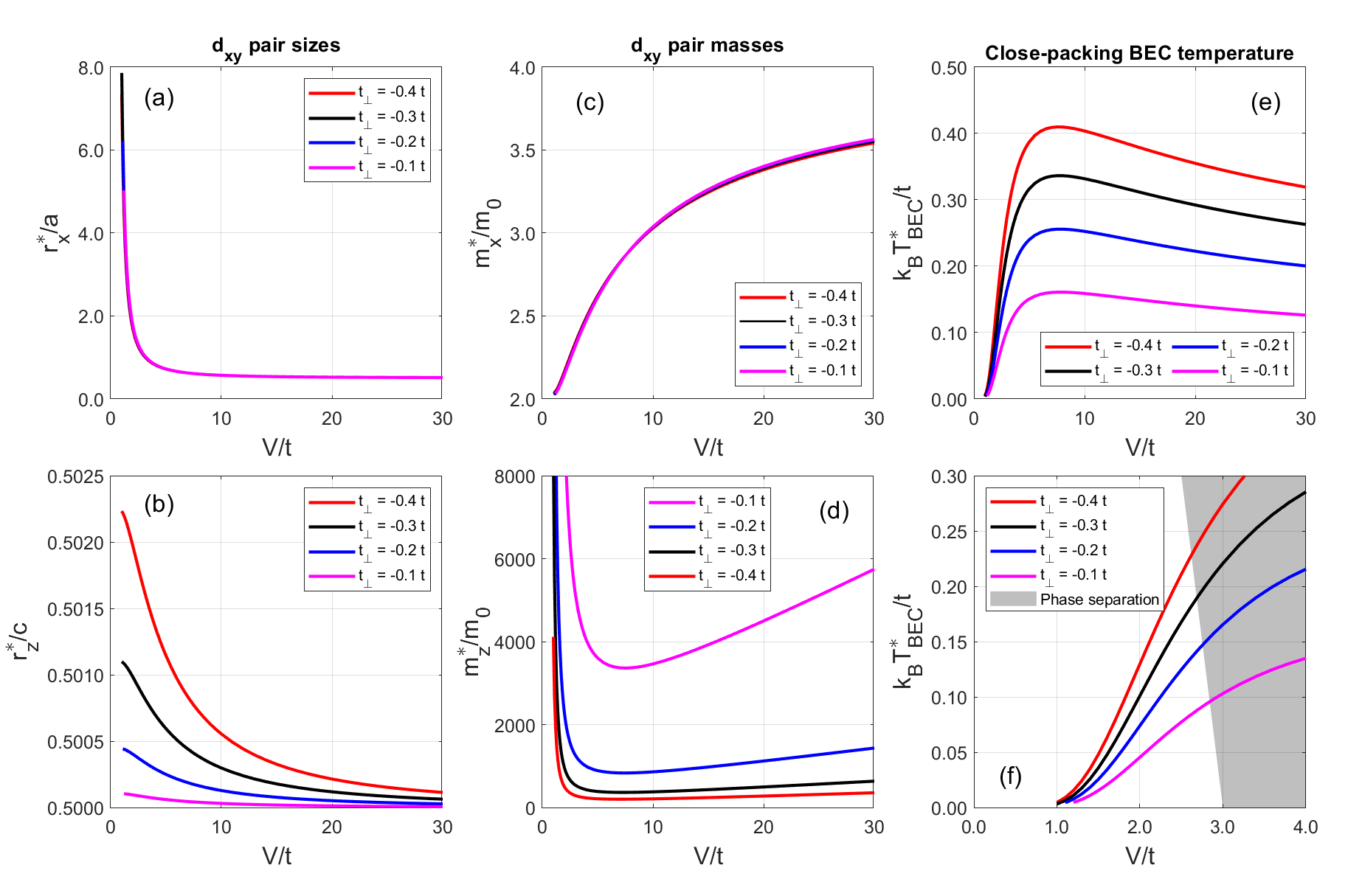}
\caption{Properties of the ground-state, $d_{xy}$-symmetric, hole pairs in the BCT $UV$ model as a function of $V$ and $t_{\perp}$. $U = 30\,t$. (a) Pair effective radius in the $xy$ plane, $r^{\ast}_{x}$, in units of lattice constant $a$. $r^{\ast}_{x}$ diverges towards threshold, $V \to 0$, and saturates at $0.5$ at strong attraction. (b) Pair effective radius across the planes, $r^{\ast}_{z}$, in units of lattice constant $c$. $r^{\ast}_{z} \approx 0.5$ at all relevant $V$'s. Both $r^{\ast}_{x}$ and $r^{\ast}_{z}$ are barely dependent on the interlayer hopping, $t_{\perp}$. (c) In-plane pair effective mass, $m^{\ast}_{x}$, in units of the $(xy)$ free mass $m_{0} = \hbar^{2}/(2ta^2)$. $m^{\ast}_{x}$ displays the {\em light-pair effect} \cite{Alexandrov2002A,Hague2007,Hague2007C,Hague2008,Adebanjo2022,Kornilovitch2023,Adebanjo2024}, and never exceeds $4 m_0$ even in the strong-attraction limit \cite{Kornilovitch2023}. $m^{\ast}_{x}$ is barely dependent on the interlayer hopping, $t_{\perp}$. (d) Out-of-plane pair effective mass, $m^{\ast}_{z}$, in units of $m_0$. The out-of-plane mass is {\em not} light, but $m^{\ast}_{z}(V)$ exhibits very unusual {\em nonmonotonic} behavior. At weak coupling, $m^{\ast}_{z}$ {\em decreases} with $V$ via the mechanism described in the main text. But at strong coupling, the mass crosses over to a second-order-in-hopping regime, $m^{\ast}_{z} \propto V/t^{2}_{\perp}$. (e) Close-packing $T^{\ast}_{\rm BEC}$, Eq.~(\ref{D:eq:eight}), in units of $t$. The $V$ dependence is non-monotonic, reflecting the similar behavior of $m^{\ast}_{z}(V)$. (f) Zoom in to the physically relevant region of small $V$. The shaded area schematically indicates the domain of clustering and phase separation \cite{Kornilovitch2022}. Note that $T^{\ast}_{\rm BEC}$ systematically increases with $V$ until phase separation takes place. The highest $T^{\ast}_{\rm BEC}$ occurs near clustering and other charge instabilities.} 
\label{D:fig:four}
\end{figure*}

\section{\label{D:sec:five}
Pair properties and close-packed $T_c$
}

Real-space su\-perconductivity is a Bose-Einstein condensation of preformed pairs \cite{Eagles1969,Alexandrov1981,Micnas1990,Alexandrov1994}. We emphasize that here we are dealing with genuine bound states that are formed without the help of a Fermi sea. Equation~(\ref{D:eq:six}) describes hole pairs with zero total momenta that eventually form a superconducting condensate. Pairs not in condensate continue to move with nonzero momenta. Their properties can also be inferred from the exact solution outlined in Appendix~\ref{D:sec:app:a} and Ref.~\cite{Kornilovitch2023}. Unpaired holes do not form a Fermi sea but behave as a non-degenerate Fermi gas. In this respect, the mechanism is similar to the bipolaronic one \cite{Alexandrov1981,Alexandrov1994}. The anisotropic BEC temperature is
\begin{equation}
k_{B} T_{\rm BEC} = \frac{ 3.3 \, \hbar^2 \, n^{2/3}_{p} }
                         { ( m^{\ast}_x \, m^{\ast}_y \, m^{\ast}_z )^{1/3} }  \: . 
\label{D:eq:seven}
\end{equation}
Continuous-space approximation is unnecessary~\cite{Kornilovitch2015} but it nicely illustrates the qualitative point to be discussed here. According to Eq.~(\ref{D:eq:seven}), $T_{\rm BEC}$ increases with pair density until pairs begin to overlap. $T_{\rm BEC}$ reaches maxi\-mum at {\em close-packing}, i.e., when pair density $n_{p}$ equals inverse pair volume $\Omega_p$~\cite{Ivanov1994,Kornilovitch2015,Zhang2022,Kornilovitch2023,Adebanjo2024}:   
\begin{equation}
k_{B} T^{\ast}_{\rm BEC} = \frac{ 3.3 \, \hbar^2 }
                         { ( m^{\ast}_x \, m^{\ast}_y \, m^{\ast}_z )^{1/3} \, \Omega^{2/3}_{p} }  \: . 
\label{D:eq:eight}
\end{equation}
To maximize $T^{\ast}_{\rm BEC}$, light and compact pairs are needed. In some models, including the {\em attractive} Hubbard model, there is a contradiction: compact pairs (small $\Omega_p$) require large $V$'s whereas low $m^{\ast}$ require small $V$'s. Remarkably, it is not the case in the BCT lattice. Let us look at the constituents of Eq.~(\ref{D:eq:eight}) as a function of $V$. Representative curves are shown in Fig.~\ref{D:fig:four}; the data were obtained by the methods outlined in Appendix~\ref{D:sec:app:a}. The in-plane pair size $r^{\ast}_{x}$, Fig.~\ref{D:fig:four}(a), diverges near the pairing threshold, i.e., at $V \to 0$, but saturates as strong coupling. The out-of-plane pair size $r^{\ast}_{z}$, Fig.~\ref{D:fig:four}(b), stays approximately constant except for very small $V$ unaccessible for numerical solution. As a result, the pair volume $\Omega_{p} \approx ( r^{\ast}_{x})^2 r_{z}$ diverges at weak coupling but monotonically decreases with $V$ and approaches unit cell volume as $V \to \infty$. Note that $\Omega_{p}$ is very weakly dependent on $t_{\perp}$. The in-plane mass $m^{\ast}_{x,y}$, Fig.~\ref{D:fig:four}(c), is weakly $V$ dependent because of the {\em light pairs} mechanism: the pair can move in the first order in $t$ while never breaking the attractive bond \cite{Alexandrov2002A,Hague2007,Hague2007C,Hague2008,Adebanjo2022,Kornilovitch2023,Adebanjo2024}. Even in the strong-attraction limit, the mass is limited to $m^{\ast}_{xy} < 4 m_0$ \cite{Kornilovitch2023}. The out-of-plane mass $m^{\ast}_{z}$, Fig.~\ref{D:fig:four}(d), shows the most interesting behavior. At weak coupling, the pair wave function is composed of partial waves close to the band minimum $q_x , q_y = \pm \pi$. According to Eq.~(\ref{D:eq:five}), for these $q_{x,y}$ the dispersion is flat in the $z$ direction. It implies that $m^{\ast}_{z}$ is infinite at $V \to 0$. As $V$ increases, the pair wave function becomes more compact in real space and spread out in momentum space. Since partial waves away from the band minimum are $z$ dispersive, the pair becomes increasingly mobile in the out-of-plane direction. This is similar to how a bound pair can become mobile in a flat-band system \cite{Torma2018}. However, at strong coupling, $m^{\ast}_{z}$ begins to grow again. The out-of-plane mass is {\em not} light in the sense that pair movement is second-order in hopping. As a result, the mass grows as $m^{\ast}_{z} \propto V/t^2_{\perp}$ at large $V$. The mass passes through a minimum at intermediate $V$. Such a behavior is highly unusual, and is one more unique feature of the model, Eq.~(\ref{D:eq:four}).

\begin{figure}[t]
\includegraphics[width=0.48\textwidth]{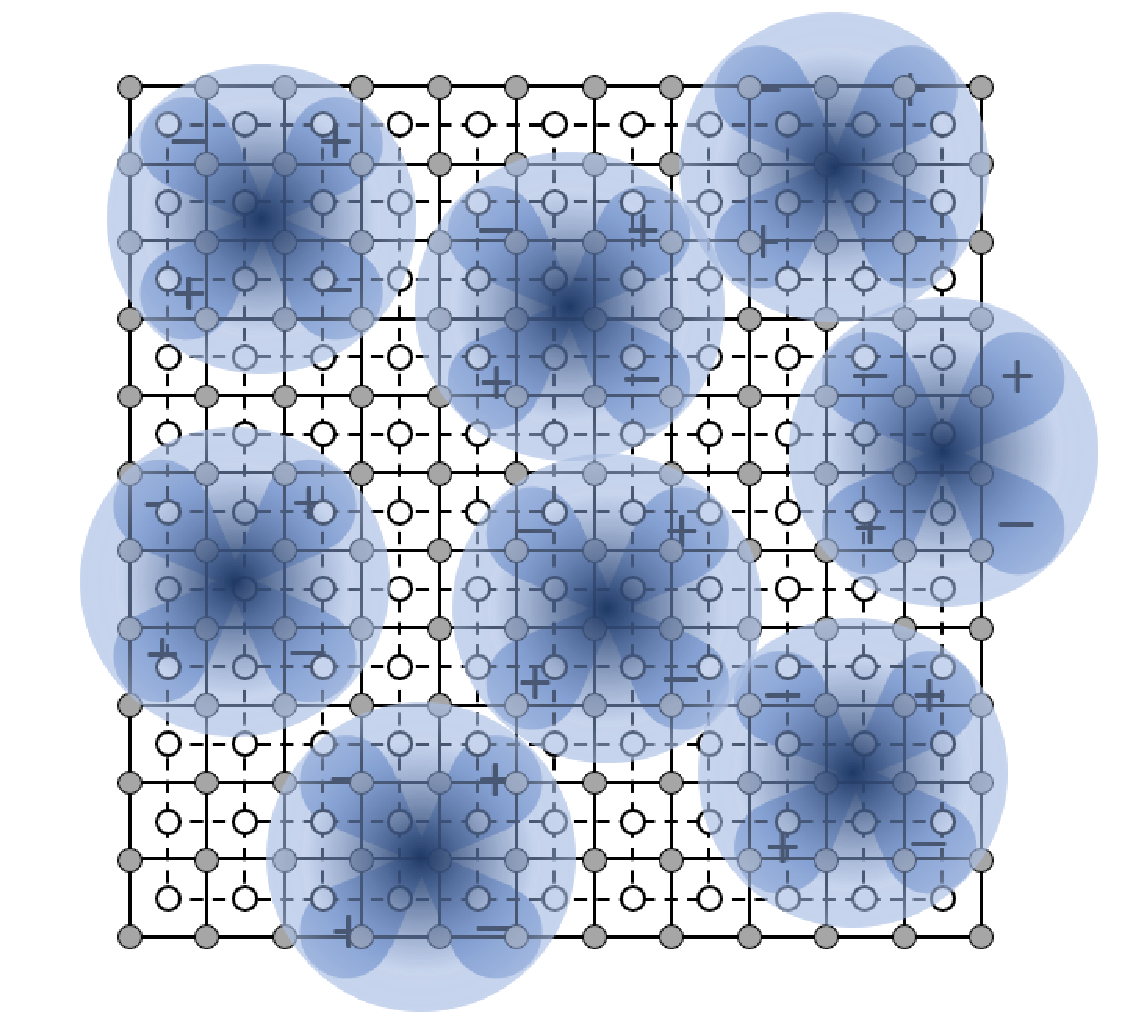}
\caption{Schematic illustration of BEC of $d$-symmetric hole pairs in the underdoped regime. {\em The coupling is still weak} in the sense that the pairs do not collapse to neighboring sites but instead occupy multiple unit cells and remain mobile. This picture roughly corresponds to a doping $x \sim 0.11$, i.e., approximately eight pairs per $12 \times 12$ unit cells {\em per layer}. Only pairs from one layer are shown to not obscure the view. With pairs' linear size of $3-4$ unit cells, the density is low enough for the pairs to not overlap and for the entire BEC picture to hold. Increasing $V$ compactifies the pairs and enables higher packing densities and hence higher $T^{\ast}_{\rm BEC}$.} 
\label{D:fig:five}
\end{figure}

Consider now the close-packing BEC temperature, Eq.~(\ref{D:eq:eight}) and Fig.~\ref{D:fig:four}(e). At weak coupling, two factors out of four in the denominator ($\Omega_p$ and $m^{\ast}_{z}$) decrease with $V$ while the other two ($m^{\ast}_{x,y}$) stay approximately constant. As a result, $T^{\ast}_{\rm BEC}$ monotonically grows with $V$. At strong coupling, the pair volume saturates and $T^{\ast}_{\rm BEC}$ begins to fall reflecting an increase in $m^{\ast}_{z}$. Thus, $T^{\ast}_{\rm BEC}$ as a function of $V$ passes through a maximum. Importantly, the optimal $V$'s are {\em larger} than the typical clustering and phase separation thresholds in $UV$ models, $V \sim (3 - 4) t$, \cite{Kornilovitch2022}. This point is further illustrated in Fig.~\ref{D:fig:four}(f) where the low-$V$ region of Fig.~\ref{D:fig:four}(e) is magnified. The domain of clustering and phase separation is schematically grayed out. Examination of the figure leads to an important conclusion: the maximal attainable $T_c$ in a given system increases with $V$ because pairs become increasingly compact, hence allowing for larger packing densities, while remaining relatively mobile. The above mechanism of boosting $T_c$ works until $V$ becomes large enough to cause phase separation \cite{Kornilovitch2022,Kornilovitch2023}. This suggests that {\em the highest critical temperatures are observed near charge-ordering instabilities}. This observation helps us to understand proliferation of charge density waves, stripes and nematic phases experimentally detected in cuprates~\cite{Tranquada1995,Chang2012,Ghiringhelli2012,Comin2015,Miao2021}.   

One should emphasize that BEC superconductivity is entirely consistent with real-space pairing and weak coupling discussed here. The BEC picture is applicable when the pair size is smaller than the average distance between pairs. Even if $V$ is small and the pair size is several unit cells (that is, relatively large), the entire picture holds as long as the pair density is low enough. This scenario is schematically illustrated in Fig.~\ref{D:fig:five}. A hole doping of $x = 0.11$ corresponds to a pair-to-pair distance of 4.5 unit cells. At the same time, using $V = 0.12$ eV, $t = 600$ K, $V/t \approx 2$, the data of Fig.~\ref{D:fig:four}(a) predict $r^{\ast}_{x} \approx 1.8$ unit cells. Thus, the low-density condition is fulfilled and the entire picture remains self-consistent. This is what may be happening in the underdoped regime of LSCO.

\section{\label{D:sec:six}
Rules of real-space superconductivity
}

Based on the above considerations, three rules of real-space superconductivity are now formulated. (i) A material must possess at least one nonzero attractive $V$ that can be orbital specific \cite{Edwards2023}. (ii) $V$ must not exceed the phase separation threshold. (iii) The highest critical temperature is achieved for those $V$, $U$, and kinetic energy terms that maximize $T^{\ast}_{\rm BEC}$ given by Eq.~(\ref{D:eq:eight}). $T_c$ differences between materials of similar crystal structures, for example La$_{2-x}$Sr$_{x}$CuO$_{4}$ and HgBa$_2$CuO$_{4+\delta}$, should be explainable by differences in attractive potential $V$. One should expect a correlation of this rule with other $T_c$ predictors \cite{Vucicevic2024}. In general, larger $V$'s  result in higher $T^{\ast}_{\rm BEC}$, which suggests a route to room-temperature superconductivity. We note that some cuprate families may possess combinations of attraction $V$ and kinetic energy $K$ that do not require an additional boost from the constant density of states near the band's bottom edge. In this regard, the zero-threshold effect discovered here may not be universal across all cuprates.

\section{\label{D:sec:seven}
Summary
}

The main goal of this work has been to show that a $d$-symmetric order parameter is naturally explainable by real-space superconductivity. All it takes is an out-of-plane attraction and hole-like in-plane hopping, both of which are present in LSCO~\cite{Zhang1991,Catlow1998}. Additionally, it is found that the body-centered tetragonal lattice (BCT), the underlying crystal structure of LSCO, is particularly conducive to real-space pairing due to specifics of its hole dispersion. The findings further support the real-space mechanism of superconductivity in cuprates.


\section*{Acknowledgement}

The author wishes to thank Mona Berciu and the anonymous referees for useful comments.



\begin{appendix}

\begin{widetext}

\section{\label{D:sec:app:a}
Solution of the two-fermion problem
}

Here we provide sufficient information to derive the basic properties of Eq.~(\ref{D:eq:four}) and Figs.~\ref{D:fig:three} and \ref{D:fig:four} of the main text. A complete analysis of the body-centered-tetragonal (BCT) $UV$ model, Eq.~(\ref{D:eq:four}), including its electron sector, is a subject for future research. Here, we limit consideration to $\Gamma$ point of the pair Brillouin zone, i.e., zero total pair momentum, ${\bf P} = 0$. The solution follows the methodology developed in Ref.~\cite{Kornilovitch2023}.

\subsection{\label{D:sec:app:aone}
Spin singlets  
}

In accordance with the general methodology~\cite{Kornilovitch2023}, we introduce four out-of-plane nearest-neighbor vectors:
\begin{equation}
{\bf b}^{\prime}_1 = \frac{1}{2} \, (   {\bf x} + {\bf y} + {\bf z} )  \: ; 
\hspace{0.5cm}
{\bf b}^{\prime}_2 = \frac{1}{2} \, (   {\bf x} - {\bf y} + {\bf z} )  \: ;  
\hspace{0.5cm}
{\bf b}^{\prime}_3 = \frac{1}{2} \, ( - {\bf x} + {\bf y} + {\bf z} )  \: ;  
\hspace{0.5cm}
{\bf b}^{\prime}_4 = \frac{1}{2} \, ( - {\bf x} - {\bf y} + {\bf z} )  \: .  
\label{D:eq:aone}
\end{equation}
The singlet pair energy $E_2$ is defined from the consistency condition of the following system of five linear equations:
\begin{eqnarray}
\Phi^{+}_{{\bf 0}           }  & = & 
     - U M^{+}_{{\bf 0}{\bf 0}            } \Phi^{+}_{{\bf 0}           }
     + V M^{+}_{{\bf 0}{\bf b}^{\prime}_1 } \Phi^{+}_{{\bf b}^{\prime}_1}
     + V M^{+}_{{\bf 0}{\bf b}^{\prime}_2 } \Phi^{+}_{{\bf b}^{\prime}_2}
     + V M^{+}_{{\bf 0}{\bf b}^{\prime}_3 } \Phi^{+}_{{\bf b}^{\prime}_3}
     + V M^{+}_{{\bf 0}{\bf b}^{\prime}_4 } \Phi^{+}_{{\bf b}^{\prime}_4}
\label{D:eq:atwo}      \\
\Phi^{+}_{{\bf b}^{\prime}_1}  & = & 
     - U M^{+}_{{\bf b}^{\prime}_1 {\bf 0}            } \Phi^{+}_{\bf 0}
     + V M^{+}_{{\bf b}^{\prime}_1 {\bf b}^{\prime}_1 } \Phi^{+}_{{\bf b}^{\prime}_1}
     + V M^{+}_{{\bf b}^{\prime}_1 {\bf b}^{\prime}_2 } \Phi^{+}_{{\bf b}^{\prime}_2}
     + V M^{+}_{{\bf b}^{\prime}_1 {\bf b}^{\prime}_3 } \Phi^{+}_{{\bf b}^{\prime}_3}
     + V M^{+}_{{\bf b}^{\prime}_1 {\bf b}^{\prime}_4 } \Phi^{+}_{{\bf b}^{\prime}_4}
\label{D:eq:athree}    \\
\Phi^{+}_{{\bf b}^{\prime}_2}  & = & 
     - U M^{+}_{{\bf b}^{\prime}_2 {\bf 0}            } \Phi^{+}_{\bf 0}
     + V M^{+}_{{\bf b}^{\prime}_2 {\bf b}^{\prime}_1 } \Phi^{+}_{{\bf b}^{\prime}_1}
     + V M^{+}_{{\bf b}^{\prime}_2 {\bf b}^{\prime}_2 } \Phi^{+}_{{\bf b}^{\prime}_2}
     + V M^{+}_{{\bf b}^{\prime}_2 {\bf b}^{\prime}_3 } \Phi^{+}_{{\bf b}^{\prime}_3}
     + V M^{+}_{{\bf b}^{\prime}_2 {\bf b}^{\prime}_4 } \Phi^{+}_{{\bf b}^{\prime}_4}
\label{D:eq:afour}     \\
\Phi^{+}_{{\bf b}^{\prime}_3}  & = & 
     - U M^{+}_{{\bf b}^{\prime}_3 {\bf 0}            } \Phi^{+}_{\bf 0}
     + V M^{+}_{{\bf b}^{\prime}_3 {\bf b}^{\prime}_1 } \Phi^{+}_{{\bf b}^{\prime}_1}
     + V M^{+}_{{\bf b}^{\prime}_3 {\bf b}^{\prime}_2 } \Phi^{+}_{{\bf b}^{\prime}_2}
     + V M^{+}_{{\bf b}^{\prime}_3 {\bf b}^{\prime}_3 } \Phi^{+}_{{\bf b}^{\prime}_3}
     + V M^{+}_{{\bf b}^{\prime}_3 {\bf b}^{\prime}_4 } \Phi^{+}_{{\bf b}^{\prime}_4}
\label{D:eq:afive}     \\
\Phi^{+}_{{\bf b}^{\prime}_4}  & = & 
     - U M^{+}_{{\bf b}^{\prime}_4 {\bf 0}            } \Phi^{+}_{\bf 0}
     + V M^{+}_{{\bf b}^{\prime}_4 {\bf b}^{\prime}_1 } \Phi^{+}_{{\bf b}^{\prime}_1}
     + V M^{+}_{{\bf b}^{\prime}_4 {\bf b}^{\prime}_2 } \Phi^{+}_{{\bf b}^{\prime}_2}
     + V M^{+}_{{\bf b}^{\prime}_4 {\bf b}^{\prime}_3 } \Phi^{+}_{{\bf b}^{\prime}_3}
     + V M^{+}_{{\bf b}^{\prime}_4 {\bf b}^{\prime}_4 } \Phi^{+}_{{\bf b}^{\prime}_4} \: . 
\label{D:eq:asix}
\end{eqnarray}
Here:
\begin{eqnarray}
M^{+}_{{\bf 0}{\bf 0}}              & = & M_{000}  \: ,
\label{D:eq:aseven}    \\
M^{+}_{{\bf 0}{\bf b}^{\prime}_{1}} = M^{+}_{{\bf 0}{\bf b}^{\prime}_{2}} =
M^{+}_{{\bf 0}{\bf b}^{\prime}_{3}} = M^{+}_{{\bf 0}{\bf b}^{\prime}_{4}}
& = & 2 M_{\frac{1}{2}\frac{1}{2}1}                \: ,
\label{D:eq:aeight}    \\
M^{+}_{{\bf b}^{\prime}_{1}{\bf 0}} = M^{+}_{{\bf b}^{\prime}_{2}{\bf 0}} =
M^{+}_{{\bf b}^{\prime}_{3}{\bf 0}} = M^{+}_{{\bf b}^{\prime}_{4}{\bf 0}}  
& = &   M_{\frac{1}{2}\frac{1}{2}1}                \: ,
\label{D:eq:anine}     \\
M^{+}_{{\bf b}^{\prime}_1 {\bf b}^{\prime}_1} =
M^{+}_{{\bf b}^{\prime}_2 {\bf b}^{\prime}_2} =
M^{+}_{{\bf b}^{\prime}_3 {\bf b}^{\prime}_3} =   
M^{+}_{{\bf b}^{\prime}_4 {\bf b}^{\prime}_4} 
& = & M_{000} + M_{112}                            \: , 
\label{D:eq:aten}      \\
M^{+}_{{\bf b}^{\prime}_1 {\bf b}^{\prime}_2} =
M^{+}_{{\bf b}^{\prime}_2 {\bf b}^{\prime}_1} =
M^{+}_{{\bf b}^{\prime}_3 {\bf b}^{\prime}_4} =
M^{+}_{{\bf b}^{\prime}_4 {\bf b}^{\prime}_3} 
& = & M_{010} + M_{102}                            \: ,
\label{D:eq:atwelve}   \\
M^{+}_{{\bf b}^{\prime}_1 {\bf b}^{\prime}_3} =
M^{+}_{{\bf b}^{\prime}_3 {\bf b}^{\prime}_1} =
M^{+}_{{\bf b}^{\prime}_2 {\bf b}^{\prime}_4} =
M^{+}_{{\bf b}^{\prime}_4 {\bf b}^{\prime}_2} 
& = & M_{100} + M_{012}                           \: ,
\label{D:eq:athirteen} \\
M^{+}_{{\bf b}^{\prime}_1 {\bf b}^{\prime}_4} =
M^{+}_{{\bf b}^{\prime}_4 {\bf b}^{\prime}_1} =
M^{+}_{{\bf b}^{\prime}_2 {\bf b}^{\prime}_3} =
M^{+}_{{\bf b}^{\prime}_3 {\bf b}^{\prime}_2} 
& = & M_{110} + M_{002}                           \: ,
\label{D:eq:afourteen} 
\end{eqnarray}
where
\begin{equation}
M_{nml} \equiv \int\limits^{\pi}_{-\pi} \!\! \int\limits^{\pi}_{-\pi}  \!\! \int\limits^{\pi}_{-\pi} 
\frac{ {\rm d} q_x {\rm d} q_y {\rm d} q_z }{ ( 2 \pi )^3 }
\frac{ \cos{n q_x} \, \cos{m q_y} \, \cos{l q_z} } 
     { | E_{2} | + 4t ( \cos{q_x} + \cos{q_y} ) + 
     16 t_{\perp} \cos{\frac{q_x}{2}} \cos{\frac{q_y}{2}} \cos{q_z} } \: . 
\label{D:eq:afifteen}
\end{equation}
Additionally, $M_{100} = M_{010}$ and $M_{102} = M_{012}$. Introducing a new basis:
\begin{eqnarray}
\Phi_{0}        & = & \Phi^{+}_{\bf 0} 
\label{D:eq:asixteen}    \\
\Phi_{s}        & = & \frac{1}{2} 
\left( \Phi^{+}_{{\bf b}^{\prime}_1} + \Phi^{+}_{{\bf b}^{\prime}_2} + 
       \Phi^{+}_{{\bf b}^{\prime}_3} + \Phi^{+}_{{\bf b}^{\prime}_4} \right) 
\label{D:eq:aseventeen}  \\
\Phi_{d_{xz}}   & = & \frac{1}{2} 
\left( \Phi^{+}_{{\bf b}^{\prime}_1} + \Phi^{+}_{{\bf b}^{\prime}_2} - 
       \Phi^{+}_{{\bf b}^{\prime}_3} - \Phi^{+}_{{\bf b}^{\prime}_4} \right) 
\label{D:eq:aeighteen}   \\
\Phi_{d_{yz}}   & = & \frac{1}{2} 
\left( \Phi^{+}_{{\bf b}^{\prime}_1} - \Phi^{+}_{{\bf b}^{\prime}_2} + 
       \Phi^{+}_{{\bf b}^{\prime}_3} - \Phi^{+}_{{\bf b}^{\prime}_4} \right) 
\label{D:eq:anineteen}   \\
\Phi_{d_{xy}}   & = & \frac{1}{2} 
\left( \Phi^{+}_{{\bf b}^{\prime}_1} - \Phi^{+}_{{\bf b}^{\prime}_2} - 
       \Phi^{+}_{{\bf b}^{\prime}_3} + \Phi^{+}_{{\bf b}^{\prime}_4} \right)  \: ,
\label{D:eq:atwenty}  
\end{eqnarray}
the system, Eqs.~(\ref{D:eq:atwo})--(\ref{D:eq:asix}), is diagonalized into the following blocks. 

\noindent
\underline{Extended $s$:} 
\begin{eqnarray}
\Phi_{0} & = & - U M_{000} \Phi_{0} + 4 V M_{\frac{1}{2}\frac{1}{2}1} \Phi_{s}
\label{D:eq:atwentyone}  \\
\Phi_{s} & = & - 2 U M_{\frac{1}{2}\frac{1}{2}1} \Phi_{0}
+ V \left( M_{000} + M_{112} + 2 M_{100} + 2 M_{102} + M_{110} + M_{002} \right) \Phi_{s} \: .
\label{D:eq:atwentytwo}  
\end{eqnarray}
\underline{Twice-degenerate doublet $(d_{xz},d_{yz})$:} 
\begin{eqnarray}
\Phi_{d_{xz}} & = & V \left( M_{000} + M_{112} - M_{110} - M_{002} \right) \Phi_{d_{xz}} 
\label{D:eq:atwentythree}  \\
\Phi_{d_{yz}} & = & V \left( M_{000} + M_{112} - M_{110} - M_{002} \right) \Phi_{d_{yz}} \: .
\label{D:eq:atwentyfour}  
\end{eqnarray}
\underline{A separate $d_{xy}$ state:} 
\begin{equation}
\Phi_{d_{xy}} = 
V \left( M_{000} + M_{112} - 2 M_{100} - 2 M_{102} + M_{110} + M_{002} \right) \Phi_{d_{xy}} \: .
\label{D:eq:atwentyfive}  
\end{equation}
The above equations define pair energies $E_{2}$ vs. $U, V, t, t_{\perp}$ for their respective symmetries. An example $E_{2}(V)$ is plotted in Fig.~\ref{D:fig:three} of the main text.

\subsection{\label{D:sec:app:atwo}
Spin triplets  
}

The general spin-triplet pair solution is described by the linear system of four equations:
\begin{eqnarray}
\Phi^{-}_{{\bf b}^{\prime}_1}  & = & 
     + V M^{-}_{{\bf b}^{\prime}_1 {\bf b}^{\prime}_1 } \Phi^{-}_{{\bf b}^{\prime}_1}
     + V M^{-}_{{\bf b}^{\prime}_1 {\bf b}^{\prime}_2 } \Phi^{-}_{{\bf b}^{\prime}_2}
     + V M^{-}_{{\bf b}^{\prime}_1 {\bf b}^{\prime}_3 } \Phi^{-}_{{\bf b}^{\prime}_3}
     + V M^{-}_{{\bf b}^{\prime}_1 {\bf b}^{\prime}_4 } \Phi^{-}_{{\bf b}^{\prime}_4} 
\label{D:eq:atwentysix}    \\
\Phi^{-}_{{\bf b}^{\prime}_2}  & = & 
     + V M^{-}_{{\bf b}^{\prime}_2 {\bf b}^{\prime}_1 } \Phi^{-}_{{\bf b}^{\prime}_1}
     + V M^{-}_{{\bf b}^{\prime}_2 {\bf b}^{\prime}_2 } \Phi^{-}_{{\bf b}^{\prime}_2}
     + V M^{-}_{{\bf b}^{\prime}_2 {\bf b}^{\prime}_3 } \Phi^{-}_{{\bf b}^{\prime}_3}
     + V M^{-}_{{\bf b}^{\prime}_2 {\bf b}^{\prime}_4 } \Phi^{-}_{{\bf b}^{\prime}_4} 
\label{D:eq:atwentyseven}     \\
\Phi^{-}_{{\bf b}^{\prime}_3}  & = & 
     + V M^{-}_{{\bf b}^{\prime}_3 {\bf b}^{\prime}_1 } \Phi^{-}_{{\bf b}^{\prime}_1}
     + V M^{-}_{{\bf b}^{\prime}_3 {\bf b}^{\prime}_2 } \Phi^{-}_{{\bf b}^{\prime}_2}
     + V M^{-}_{{\bf b}^{\prime}_3 {\bf b}^{\prime}_3 } \Phi^{-}_{{\bf b}^{\prime}_3}
     + V M^{-}_{{\bf b}^{\prime}_3 {\bf b}^{\prime}_4 } \Phi^{-}_{{\bf b}^{\prime}_4} 
\label{D:eq:atwentyeight}     \\
\Phi^{-}_{{\bf b}^{\prime}_4}  & = & 
     + V M^{-}_{{\bf b}^{\prime}_4 {\bf b}^{\prime}_1 } \Phi^{-}_{{\bf b}^{\prime}_1}
     + V M^{-}_{{\bf b}^{\prime}_4 {\bf b}^{\prime}_2 } \Phi^{-}_{{\bf b}^{\prime}_2}
     + V M^{-}_{{\bf b}^{\prime}_4 {\bf b}^{\prime}_3 } \Phi^{-}_{{\bf b}^{\prime}_3}
     + V M^{-}_{{\bf b}^{\prime}_4 {\bf b}^{\prime}_4 } \Phi^{-}_{{\bf b}^{\prime}_4} \: . 
\label{D:eq:atwentynine}
\end{eqnarray}
Here
\begin{eqnarray}
M^{-}_{{\bf b}^{\prime}_1 {\bf b}^{\prime}_1} =
M^{-}_{{\bf b}^{\prime}_2 {\bf b}^{\prime}_2} =
M^{-}_{{\bf b}^{\prime}_3 {\bf b}^{\prime}_3} =   
M^{-}_{{\bf b}^{\prime}_4 {\bf b}^{\prime}_4} 
& = & M_{000} - M_{112}                            \: , 
\label{D:eq:athirty}      \\
M^{-}_{{\bf b}^{\prime}_1 {\bf b}^{\prime}_2} =
M^{-}_{{\bf b}^{\prime}_2 {\bf b}^{\prime}_1} =
M^{-}_{{\bf b}^{\prime}_3 {\bf b}^{\prime}_4} =
M^{-}_{{\bf b}^{\prime}_4 {\bf b}^{\prime}_3} 
& = & M_{010} - M_{102}                            \: ,
\label{D:eq:athirtyone}   \\
M^{-}_{{\bf b}^{\prime}_1 {\bf b}^{\prime}_3} =
M^{-}_{{\bf b}^{\prime}_3 {\bf b}^{\prime}_1} =
M^{-}_{{\bf b}^{\prime}_2 {\bf b}^{\prime}_4} =
M^{-}_{{\bf b}^{\prime}_4 {\bf b}^{\prime}_2} 
& = & M_{100} - M_{012}                            \: ,
\label{D:eq:athirtytwo} \\
M^{+}_{{\bf b}^{\prime}_1 {\bf b}^{\prime}_4} =
M^{+}_{{\bf b}^{\prime}_4 {\bf b}^{\prime}_1} =
M^{+}_{{\bf b}^{\prime}_2 {\bf b}^{\prime}_3} =
M^{+}_{{\bf b}^{\prime}_3 {\bf b}^{\prime}_2} 
& = & M_{110} - M_{002}                            \: .
\label{D:eq:athirtythree} 
\end{eqnarray}
Additionally, $M_{100} = M_{010}$ and $M_{102} = M_{012}$. Introducing a new basis:
\begin{eqnarray}
\Phi_{p_x}   & = & \frac{1}{2} 
\left( \Phi^{-}_{{\bf b}^{\prime}_1} + \Phi^{-}_{{\bf b}^{\prime}_2} - 
       \Phi^{-}_{{\bf b}^{\prime}_3} - \Phi^{-}_{{\bf b}^{\prime}_4} \right) 
\label{D:eq:athirtyfour}  \\
\Phi_{p_y}   & = & \frac{1}{2} 
\left( \Phi^{-}_{{\bf b}^{\prime}_1} - \Phi^{-}_{{\bf b}^{\prime}_2} + 
       \Phi^{-}_{{\bf b}^{\prime}_3} - \Phi^{-}_{{\bf b}^{\prime}_4} \right) 
\label{D:eq:athirtyfive}   \\
\Phi_{p_z}   & = & \frac{1}{2} 
\left( \Phi^{-}_{{\bf b}^{\prime}_1} + \Phi^{-}_{{\bf b}^{\prime}_2} + 
       \Phi^{-}_{{\bf b}^{\prime}_3} + \Phi^{-}_{{\bf b}^{\prime}_4} \right) 
\label{D:eq:athirtysix}   \\
\Phi_{f}     & = & \frac{1}{2} 
\left( \Phi^{-}_{{\bf b}^{\prime}_1} - \Phi^{-}_{{\bf b}^{\prime}_2} - 
       \Phi^{-}_{{\bf b}^{\prime}_3} + \Phi^{-}_{{\bf b}^{\prime}_4} \right)  \: ,
\label{D:eq:athirtyseven}  
\end{eqnarray}
the system, Eqs.~(\ref{D:eq:atwentysix})--(\ref{D:eq:atwentynine}), is diagonalized into the following blocks. 

\noindent
\underline{Twice-degenerate doublet $(p_{x},p_{y})$:} 
\begin{eqnarray}
\Phi_{p_{x}} & = & V \left( M_{000} - M_{112} - M_{110} + M_{002} \right) \Phi_{p_{x}} 
\label{D:eq:athirtyeight}  \\
\Phi_{p_{y}} & = & V \left( M_{000} - M_{112} - M_{110} + M_{002} \right) \Phi_{p_{y}} \: .
\label{D:eq:athirtynine}  
\end{eqnarray}
\underline{A separate $p_{z}$ state:} 
\begin{equation}
\Phi_{p_{z}} = 
V \left( M_{000} - M_{112} + 2 M_{100} - 2 M_{102} + M_{110} - M_{002} \right) \Phi_{p_{z}} \: .
\label{D:eq:aforty}  
\end{equation}
\underline{A separate $f$ state:} 
\begin{equation}
\Phi_{f} = 
V \left( M_{000} - M_{112} - 2 M_{100} + 2 M_{102} + M_{110} - M_{002} \right) \Phi_{f} \: .
\label{D:eq:afortyone}  
\end{equation}
The above equations define triplet pair energies $E_{2}$ vs. $U, V, t, t_{\perp}$ for their respective symmetries. An example $E_{2}(V)$ is plotted in Fig.~\ref{D:fig:three} of the main text.

\subsection{\label{D:sec:app:athree}
Pair radius  
}

In this section, an efficient method of calculating pair radius is described. Consideration is limited to spin singlets at ${\bf P} = 0$. The effective radius is defined as
\begin{equation}
\left( r^{\ast}_{\alpha} \right)^2 \equiv \langle m^{2}_{\alpha} \rangle = 
\frac{\sum_{\bf m} m^2_{\alpha} \psi^{+\ast}( {\bf m}, {\bf 0} ) \psi^{+}( {\bf m}, {\bf 0} ) }
     {\sum_{\bf m}              \psi^{+\ast}( {\bf m}, {\bf 0} ) \psi^{+}( {\bf m}, {\bf 0} ) }
\equiv \frac{J_{\alpha}}{J_0} \: .
\label{D:eq:cone}  
\end{equation}
Here, $\psi^{+}$ is the real-space wave function \cite{Kornilovitch2023}
\begin{equation}
\psi^{+}( {\bf m} , {\bf 0} ) = \frac{1}{2N} \sum_{\bf k} 
\left( e^{ i {\bf k} {\bf m} } + e^{ - i {\bf k} {\bf m} } \right) \phi^{+}_{ {\bf k}, -{\bf k} } \: .
\label{D:eq:ctwo}  
\end{equation}
The momentum-space wave function is expressible via $\Phi^{+}_{\bf b}$ \cite{Kornilovitch2023}
\begin{equation}
\phi^{+}_{ {\bf k}_1 , {\bf k}_2 } = 
\frac{U}{ E - \varepsilon_{{\bf k}_1} - \varepsilon_{{\bf k}_1} } \, \Phi^{+}_{\bf 0}({\bf k}_1 + {\bf k}_2)
+ \sum_{{\bf b}^{\prime}} V_{{\bf b}^{\prime}} \: 
\frac{ e^{ - {\bf k}_1 {\bf b}^{\prime} } + e^{ - {\bf k}_2 {\bf b}^{\prime} } }
{ E - \varepsilon_{{\bf k}_1} - \varepsilon_{{\bf k}_1} } \, \Phi^{+}_{{\bf b}^{\prime}}({\bf k}_1 + {\bf k}_2) \: .
\label{D:eq:cthree}  
\end{equation}
In our problem, $V_{{\bf b}^{\prime}} = -V$ for all four ${\bf b}^{\prime}$ defined in Eq.~(\ref{D:eq:aone}).  Substitution of Eq.~(\ref{D:eq:ctwo}) into the denominator of Eq.~(\ref{D:eq:cone}) yields
\begin{equation}
J_0 = \frac{1}{N} \sum_{\bf k} \phi^{+ \ast}_{ {\bf k}, -{\bf k} } \, \phi^{+}_{ {\bf k}, -{\bf k} } =
\frac{1}{N} \sum_{\bf k} \vert \phi^{+}_{ {\bf k}, -{\bf k} } \vert^2 \: .
\label{D:eq:cfour}  
\end{equation}
In the numerator of Eq.~(\ref{D:eq:cone}), $m_{\alpha}$ can be represented as a derivative of $e^{\pm i {\bf k} {\bf m}}$. After transformations, and utilizing $\phi^{+}_{{\bf k}_1 {\bf k}_2} = \phi^{+}_{{\bf k}_2 {\bf k}_1}$, one obtains 
\begin{equation}
J_{\alpha} = \frac{1}{N^2} \sum_{\bf m} \left( \sum_{\bf k} \phi^{+}_{ {\bf k}, -{\bf k} } 
\frac{\partial}{\partial k_{\alpha}} \, e^{ i {\bf k} {\bf m} }  \right) 
\left( \sum_{{\bf k}^{\prime}} \phi^{+ \ast}_{ {\bf k}^{\prime} , -{\bf k}^{\prime} } 
\frac{\partial}{\partial k^{\prime}_{\alpha}} \, e^{ - i {\bf k}^{\prime} {\bf m} } \right) .
\label{D:eq:cfive}  
\end{equation}
One can apply the Green's theorem for periodic functions to both sums over ${\bf k}$ \cite{Ashcroft1976}. Integration by parts leads to 
\begin{equation}
J_{\alpha} = \frac{1}{N} \sum_{\bf k}  
\frac{\partial}{\partial k_{\alpha}} \left( \phi^{+}_{ {\bf k}, -{\bf k} } \right)
\frac{\partial}{\partial k_{\alpha}} \left( \phi^{+\ast}_{ {\bf k}, -{\bf k} } \right) .
\label{D:eq:csix}  
\end{equation}
Utilizing Eq.~(\ref{D:eq:cthree}) for ${\bf k}_1 + {\bf k}_2 = {\bf P} = 0$, one obtains
\begin{equation}
\frac{\partial}{\partial k_{\alpha}} \left( \phi^{+}_{ {\bf k}, -{\bf k} } \right) = 
\left[ \frac{ 2 U }{ ( E - 2 \varepsilon_{\bf k} )^2 } 
\frac{ \partial \varepsilon_{\bf k} }{ \partial k_{\alpha} } \right] \Phi^{+}_{\bf 0}({\bf 0})
 + \sum_{{\bf b}^{\prime}} V_{{\bf b}^{\prime}} \left[ 
 - \frac{ 2 b_{\alpha} \sin{({\bf k}{\bf b}^{\prime})}}{ E - 2 \varepsilon_{\bf k} }
 + \frac{ 4 \cos{({\bf k}{\bf b}^{\prime})} }{ ( E - 2 \varepsilon_{\bf k} )^2 } 
\frac{ \partial \varepsilon_{\bf k} }{ \partial k_{\alpha} }
 \right] \Phi^{+}_{{\bf b}^{\prime}}({\bf 0}) .
\label{D:eq:cseven}  
\end{equation}
The second factor in Eq.~(\ref{D:eq:csix}) is a complex conjugate of Eq.~(\ref{D:eq:cseven}). All together, Eqs.~(\ref{D:eq:cone}), (\ref{D:eq:cthree}), (\ref{D:eq:cfour}), (\ref{D:eq:csix}), and (\ref{D:eq:cseven}) provide a general recipe of calculating pair radius. Once pair energy $E$ and eigenvector $\{ \Phi^{+}_{\bf 0} , \Phi^{+}_{\bf b} \}$ are known from the main linear system, $r_{\alpha}$ only requires calculation of two momentum integrals. 

In some situations, the general procedure simplifies significantly. As an example, consider $d_{xy}$ pair state whose energy is defined by Eq.~(\ref{D:eq:atwentyfive}). Inverting the basis transformation, Eqs.~(\ref{D:eq:aseventeen})--(\ref{D:eq:atwenty}), one obtains
\begin{equation}
\Phi^{+}_{\bf 0} = 0 \: ; \hspace{1.0cm}
\Phi^{+}_{{\bf b}^{\prime}_1} = - \Phi^{+}_{{\bf b}^{\prime}_2} = 
- \Phi^{+}_{{\bf b}^{\prime}_3} = \Phi^{+}_{{\bf b}^{\prime}_4} 
= \frac{1}{2} \, \Phi_{d_{xy}} \: .
\label{D:eq:ceight}  
\end{equation}
The momentum-space wave function now follows from Eq.~(\ref{D:eq:cthree}):
\begin{eqnarray}
\phi^{+}_{ {\bf k}, - {\bf k} ; d_{xy} } & = & - |V| \frac{2}{ E - 2\varepsilon_{\bf k} } \left[ 
\cos{({\bf k}{\bf b}^{\prime}_1)} - \cos{({\bf k}{\bf b}^{\prime}_2)} 
- \cos{({\bf k}{\bf b}^{\prime}_3)} + \cos{({\bf k}{\bf b}^{\prime}_4)} 
\right] \frac{1}{2} \Phi_{d_{xy}} 
\nonumber \\
& = & 4 |V| \Phi_{d_{xy}} \, 
\frac{ \sin{\frac{k_x a}{2}} \sin{\frac{k_y a}{2}} \sin{\frac{k_z c}{2}} }{ E - 2 \varepsilon_{\bf k} } \: .
\label{D:eq:cnine}  
\end{eqnarray}
The constant factor $( 4 |V| \Phi_{d_{xy}} )$ drops out of the ratio of two integrals and can be disregarded. Thus, only pair energy $E_{2}$ is needed for radius calculation in this case. Utilizing this expression for $\phi^{+}_{{\bf k},-{\bf k}}$, the derivatives needed in Eq.~(\ref{D:eq:csix}) can be explicitly derived. Then everything is substituted in Eqs.~(\ref{D:eq:cfour}) and (\ref{D:eq:csix}) for numerical integration.

\section{\label{D:sec:app:b}
Bogoliubov's argument \cite{Bogoliubov1970}
}

In this section we outline the relation between macroscopic order parameter $\Delta$ and real-space pair wave function $\psi$ derived by Bogoliubov~\cite{Bogoliubov1970}. The four-point equal-time correlation function 
\begin{equation}
F( {\bf r}_1 , {\bf r}_2 ; {\bf r}^{\prime}_1 , {\bf r}^{\prime}_2 ) 
= \langle c^{\dagger}_{{\bf r}_1} c^{\dagger}_{{\bf r}_2} 
              c_{{\bf r}^{\prime}_2}  c_{{\bf r}^{\prime}_1}  \rangle      
= \sum_{\nu} N_{\nu} \Psi^{\ast}_{\nu}( {\bf r}_1 , {\bf r}_2 ) 
                         \Psi_{\nu}( {\bf r}^{\prime}_1 , {\bf r}^{\prime}_2 ) \: ,
\label{D:eq:one}
\end{equation}   
can be expanded in a complete set of normalized solutions of a two-body Schr\"odinger equation $\Psi_{\nu}$. Quantum number $\nu$ includes pair's total momentum and a discrete index of relative motion of the two particles forming the pair. In Eq.~(\ref{D:eq:one}), $c$ are fermion operators, and $\langle \ldots \rangle$ is understood as quasi-averaging, i.e., quantum-statistical averaging is performed with a fixed phase angle of the superconducting condensate and averaging over the phase angle is not performed. If $N$ is the total number of particles then $N ( N - 1 )$ is the total number of pairs. By setting ${\bf r}^{\prime}_1 = {\bf r}_1$, ${\bf r}^{\prime}_2 = {\bf r}_2$ and integrating over ${\bf r}_{1,2}$, it follows from Eq.~(\ref{D:eq:one}) that $\langle N ( N - 1 ) \rangle = \sum_{\nu} N_{\nu}$ \cite{Bogoliubov1970}. Thus, coefficient $N_{\nu}$ is the average number of pairs in a state with wave function $\Psi_{\nu}$. To describe superconducting state, one assumes a condensate of pairs in the ground state, i.e., $N_0$ is macroscopically large.    
   
Next, the four coordinates in Eq.~(\ref{D:eq:one}) are combined in two groups, $\{ {\bf r}_1 , {\bf r}_2 \}$ and $\{ {\bf r}^{\prime}_1 , {\bf r}^{\prime}_2 \}$, and are infinitely separated. The four-point correlation function splits into a product of two-point correlation functions: 
\begin{equation}
\langle c^{\dagger}_{{\bf r}_1} c^{\dagger}_{{\bf r}_2} 
        c_{{\bf r}^{\prime}_2}  c_{{\bf r}^{\prime}_1}  \rangle = 
\langle c^{\dagger}_{{\bf r}_1} c^{\dagger}_{{\bf r}_2} \rangle 
\langle c_{{\bf r}^{\prime}_2}  c_{{\bf r}^{\prime}_1}  \rangle \equiv 
\Delta^{\!\ast}( {\bf r}_1 , {\bf r}_2 ) \Delta( {\bf r}^{\prime}_1 , {\bf r}^{\prime}_2 ) \, .
\label{D:eq:two}
\end{equation}
Because of quasi-averaging, $\Delta \neq 0$. The right-hand side of Eq.~(\ref{D:eq:one}) receives contributions from unpaired $\Psi_{\nu}$ as well as from paired $\Psi_{\nu}$ not in the condensate. Bogoliubov showed that in the dilute limit, the main contribution comes from the condensate pair state with a relative wave function $\psi_{0}( {\bf r}_1 - {\bf r}_2 )$. The final result is 
\begin{equation}
\Delta( {\bf r}_1 , {\bf r}_2 ) = \sqrt{\frac{N_0}{\Omega}} \, \psi_0( {\bf r}_1 - {\bf r}_2 ) \, , 
\label{D:eq:three}
\end{equation}
where $\Omega$ is system's volume and $\psi_0(\bf r)$ is normalized to 1. This formula has an important practical implication: {\em orbital symmetry of the superconducting order parameter is the same as that of the pair wave function}. The latter can be deduced from a two-body lattice Schr\"odinger equation, which is exactly solvable~\cite{Kornilovitch2023}. In the main text, this is done for the body-centered tetragonal lattice.

\end{widetext}

\end{appendix}

\end{document}